# rECGnition_v2.0: Self-Attentive Canonical Fusion of ECG and Patient Data using deep learning for effective Cardiac Diagnostics

**Shreya Srivastava**
*Indian Institute of Technology Roorkee, Roorkee, India*
ssrivastava@bt.iitr.ac.in

**Durgesh Kumar**
*Indian Institute of Technology Roorkee, Roorkee, India*
dkumar@ce.iitr.ac.in

**Ram Jiwari**
*Indian Institute of Technology Roorkee, Roorkee, India*
ram.jiwari@ma.iitr.ac.in

**Sandeep Seth**
*All India Institute of Medical Sciences, New Delhi, India*
drsandeepseth@hotmail.com

**Deepak Sharma**
*Indian Institute of Technology Roorkee, Roorkee, India*
deepak.sharma@bt.iitr.ac.in

## ABSTRACT

***Background and Objective***: The variability in ECG readings influenced by individual patient characteristics has posed a considerable challenge to adopting automated ECG analysis in clinical settings. A novel feature fusion technique termed SACC (Self Attentive Canonical Correlation) was proposed to address this. It calibrates the ECG features based on the patient characteristics features to effectively discard benign changes in ECG morphology that arises due to patient characteristics. The overall objective of the study was to create a fast and accurate model by effectively considering the influence of patient characteristic features onto ECG morphology. The proposed novel fusion technique is combined with DPN (Dual Pathway Network) and depth-wise separable convolution to create a robust, interpretable, and fast end-to-end arrhythmia classification model named rECGnition_v2.0 (robust ECG abnormality detection v2). ***Methodology***: This study uses MIT-BIH, INCARTDB and EDB dataset to test the efficiency of rECGnition_v2.0 for various classes (5/8/10/AAMI) of arrhythmia classification. The heartbeats were segmented from given datasets and converted to 2D images (like existing methodologies) to perform its benchmarking. To investigate the influence of constituting model components, various ablation studies were performed, *i.e*. simple concatenation, CCA and proposed SACC were compared, while the importance of global and local ECG features were tested using DPN rECGnition_v2.0 model and vice versa. rECGnition_v2.0 was also benchmarked with state-of-the-art CNN models (MobileNet, EfficientNetB0, ResNet50) for overall accuracy *vs* model parameters, FLOPs, memory requirements, and prediction time. Furthermore, the inner working of the model was interpreted by comparing the activation locations in ECG before and after the SACC layer. Also, the linear separability of final feature map induced by SACC was interpreted using principal component analysis. ***Results***: The proposed architecture rECGnition_v2.0 showed a remarkable accuracy of 98.07% and an F1-score of 98.05% for classifying ten distinct classes of arrhythmia with just 82.7M FLOPs per sample, thereby going beyond the performance metrics of current state-of-the-art (SOTA) models by utilizing MIT-BIH Arrhythmia dataset. Similarly, on INCARTDB and EDB datasets, excellent F1-scores of 98.01% and 96.21% respectively was achieved for AAMI classification. ***Conclusions***: The compact architectural footprint of the rECGnition_v2.0, characterized by its lesser trainable parameters and diminished computational demands, unfurled triple advantages. Firstly, it enabled expeditious model training and real-time inference, essential for prompt diagnosis and intervention in a clinical environment. Secondly, it brings about a more interpretable model that aligns well with the requisites of the medical domain, where model interpretability is of paramount significance for clinicians' trust and broader acceptance. Thirdly, rECGnition_v2.0 is a dynamic model that can be scaled depending on the availability of patient characteristics and complexity of existing correlation.

# HIGHLIGHTS

- Dual Pathway Network (DPN) coupled with Self-Attentive Canonical Correlation (SACC) Layer, for robust arrhythmia classification.
- A novel Self-Attentive Canonical Correlation to explicitly learn the influence of external factors on the ECG morphology.
- Achievement of a significant accuracy and F1-score of 98.07% and 98.05%, respectively, in classifying ten distinct classes of arrhythmia on the MIT-BIH Arrhythmia dataset with a considerably lower computational requirement of 82.7M FLOPs per sample.
- Implementation of a compact model characterized by reduced trainable parameters and computational demands, facilitating swift model training, real-time inference, and enhanced model interpretability.

# GRAPHICAL ABSTRACT

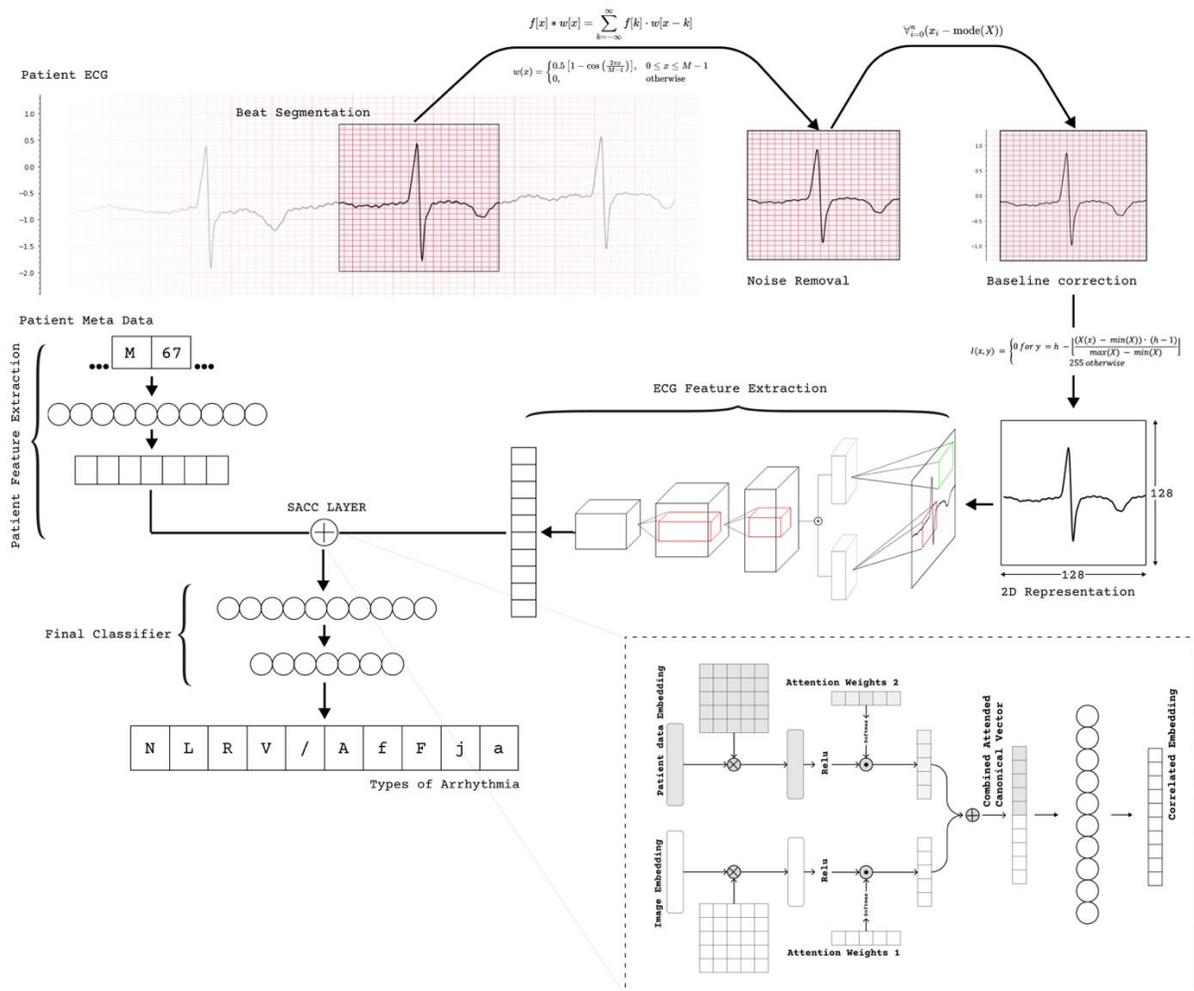

# 1. Introduction

Cardiac disorders remain a predominant health adversity with a global footprint [1, 2], requiring improved diagnostics for early diagnosis and treatment. Arrhythmias, among other cardiac problems, are hidden and dangerous if undiagnosed [3]. The electrocardiogram (ECG) is a key cardiac diagnostic procedure that shows the heart's electrical activity non-invasively [4]. ECG signals are useful for automated arrhythmia classification because their morphological details typically indicate heart health [5, 6].

With the growing popularity of deep learning, the prospect of automated cardiac arrhythmia classification has transitioned from an academic speculation to a realistic goal [7, 8]. In search of better diagnostic methods, sophisticated computational frameworks that can detect minor but crucial ECG irregularities have been explored [9, 10]. With the advent of wearable ECG collection devices that also have access to a patient's broader health data, there is a thriving interest in exploring robust methodologies to correlate patient demographic information with ECG data for faster diagnosis and detection of cardiovascular diseases at an individual level [11-13]. Research has elucidated that patient demographics such as age, gender, and ethnicity can significantly influence ECG morphology [14-16]. The morphological variations of ECG signals, often intertwined with demographic attributes, play a pivotal role in accurately detecting and interpreting cardiovascular anomalies. Modern wearable devices can continuously monitor a wide range of health parameters along with ECG. This makes it possible to combine demographic and ECG data in a unique way [17, 18]. This fusion, enabled by advanced computational models like Self-Attentive Canonical Correlation (SACC), promises to help a more complex understanding of cardiovascular health tailored to individual patient profiles. The proposed Dual Pathway Network (DPN) and SACC are innovative arrhythmia classification methods for complicated ECG data. The SACC helps separate patients' meta-data influences, such as

| | | | |
|---:|---|---:|---|
| **/:** | Paced Beat | **DPN:** | Dual Pathway Network; |
| **a:** | Aberrated atrial premature beat | **ECG:** | Electrocardiogram; |
| **A:** | Atrial Premature beat | **E$_m$:** | ECG morphology |
| **AAMI:** | Association for advancement of medical instrumentation | **f:** | Fusion of paced and normal beat |
| **CVD:** | Cardiovascular diseases | **F:** | Fusion of ventricular and normal beat |
| **FLOPs:** | Floating Point Operations | **FP:** | False Positive |
| **FN:** | False Negative | **j:** | Nodal escape beat |
| **INCARTDB:** | St. Petersburg INCART 12-lead Arrhythmia Database | **LBBB:** | Left Bundle Branch Block beat |
| **EDB:** | European ST-T Database | **MITDB:** | MIT BIH Arrhythmia dataset |
| **N:** | Normal beat | **P$_c$:** | Patient characteristics |
| **RBBB:** | Right Bundle Branch Block beat | **SACC:** | Self-Attentive Canonical Correlation |
| **TN:** | True Negative | **TP:** | True Positive |
| **V:** | Premature Ventricular Contraction beat | **rECGnition_v2.0** | robust ECG abnormality detection v2 |

age and gender, on ECG morphology. Meanwhile, the DPN was adapted to emphasize the diagnostically crucial PQRST complex and focus on localized and global aspects of it. Due to the model's tendency to maximize the correlation for better prediction, it could discard non-arrhythmia changes more effectively with fewer parameters, thus reducing the computational cost. The compact architecture of the rECGnition_v2.0 (robust ECG abnormality detection), characterized by its reduced computational demands and lesser trainable parameters, aligns well with the needs of the real-world for swift model training, real-time inference, and enhanced model interpretability. This makes the proposed method more likely to be used in clinical settings.Using these new algorithmic combinations is expected to greatly expand the range of methods used for heart diagnosis. The potency and dominance of the proposed rECGnition_v2.0 were established against the SOTA models through rigorous experimentation on the MIT-BIH Arrhythmia dataset (MITBIH).

This research work's primary contributions are summarized below:

- Development and application of a novel architectural paradigm, DPN coupled with SACC, for robust arrhythmia classification by analysing local and global ECG heartbeat morphology features and understanding how external demographic factors affect them.
- Classifying ten arrhythmia classes on the MIT-BIH Arrhythmia dataset with 98.07% accuracy and 98.05% F1-score with 82.7m FLOPs per sample, outperforming state-of-the-art (SOTA) models in efficiency and accuracy.
- Implementing a compact model with decreased trainable parameters and computing needs allows fast model training, real-time inference, and improved model interpretability, which is essential for clinical applicability and medical domain acceptability.
- Empirical demonstration of the proposed architecture's efficacy on a benchmark dataset, advancing algorithmic frameworks for robust, efficient, and clinically deployable cardiac arrhythmia classification models, which will improve personalized cardiac healthcare and precise cardiovascular disease management by fusing demographic and ECG data.

## 2. Related work

ECG arrhythmia classification research has grown, especially since deep learning. Excellent work employed a deep neural network to classify 12 rhythm classifications using single-lead ECGs from 53,000 people approximately, indicating deep learning's potential for cardiologist-level arrhythmia classification [19]. Another important study used deep learning algorithms to analyze a large heartbeat dataset of 1,09,446 beats, divided into five kinds, revealing a pre-processing method that greatly improved ECG classification accuracy [20]. Recently, a method combining Recurrent plot with CNN was also proposed [21]. Moreover, an arrhythmia classification system utilizing a multi-head self-attention method optimises the extensive semantic information in the ECG signal, strengthening categorization [22]. Multimodal fusion was also highlighted, as existing machine learning algorithms either use manually derived characteristics or the 1D ECG input directly. Intelligent multimodal fusion could improve arrhythmia classification [23]. Interestingly, an automated classification method employing a complete ECG database inter-patient paradigm separation was created to find minority arrhythmical classes without feature extraction was developed, demonstrating the perfect example of patient metadata [24].

New ECG-based methods for inter-patient and patient-specific arrhythmia classification present a variety of solutions to the inherent problems. In particular, a study examined an automatic classification system for inter-patient ECG arrhythmia heartbeat categorization in accordance with the ANSI/AAMI EC57:1998 standard, which classifies each heartbeat as normal (*N*), SVEB (*S*), VEB (*V*), fusion (*F*), or unclassifiable (*Q*) [25]. Another novel approach suggested a single-path hybrid model with a dual attention mechanism to reduce class imbalance in automatic arrhythmia classification in inter-patient scenarios [26]. A distinct approach employing a sequence-to-sequence deep learning methodology for inter- and intra-patient ECG heartbeat categorization for arrhythmia presented a novel perspective [27]. The challenge of imbalanced datasets and the proposal of a solution to enhance cardiac arrhythmia classification methods' performance in different heart conditions, both inter and intra-patient, were the main points of another research [28]. Finally, a major study described a fully automated method for arrhythmia classification from ECG data in an inter-patient context, dividing the process into four crucial steps: pre-processing of ECG signals, heartbeat segmentation, extraction of features, and classification [29]. Patient-specific techniques can provide tailored care and better prediction, but they cannot explain why two patients with the same heart anomaly have different results. Before clinical use, patient-specific models must be trained/fine-tuned, and if there is no data, a previously trained model based on the patient's features should be picked. An intra-patient, patient information fusion model is better than a patient-specific model since it can personalize diagnoses for all patients.

## 3. Materials and Methods

### 3.1. Convolution and Depth-wise separable convolution

In convolutional neural networks, standard convolution operations involve applying a set of filters, each with dimensions $K \times K \times C$, where $K$ is the kernel size and $C$ is the number of input channels, across the input tensor to produce an output feature map. Mathematically, the output $Y$ at a spatial location $(i, j)$ for the $k$-th filter can be expressed as:

$Y_{i,j,k} = \sum_{m=1}^{K} \sum_{n=1}^{K} \sum_{c=1}^{C} X_{i+m,j+n,c} \cdot W_{m,n,c,k}$,

where $X$ is the input tensor and $W$ are the filter weights. Depth-wise separable convolutions were applied here to reduce the computational cost and model size; it decomposes the standard convolution into two parts: depth-wise convolution and pointwise convolution. In depth-wise convolution, a single filter is applied per input channel, yielding intermediate output $Z_{i,j,c} = \sum_{m=1}^{K} \sum_{n=1}^{K} X_{i+m,j+n,c} \cdot W_{m,n,c}$. This is followed by a pointwise convolution, which involves $1 \times 1$ convolutions to combine the outputs from the depth-wise step, producing the final output $Y_{i,j,k} = \sum_{c=1}^{C} Z_{i,j,c} \cdot$ $W_{1,1,c,k}$. This approach significantly decreases the number of parameters and computational complexity.

### 3.2. Dual Pathway Network (DPN)

ECG heartbeat is composed of various sinusoidal waves of varying frequency [30], so it is crucial to analyse it on various time scales. Hence, CNN blocks in our network employ a varying size kernel to extract features. The DPN architecture was implemented as a CNN that umtilizes two parallel branches. The first branch focused onextracting local, high-resolution features using a smaller receptive field. Conversely, the second branch captures the global, low-resolution features using a larger receptive field. The concatenated feature map is passed down the network for further feature extraction (Figure 1).

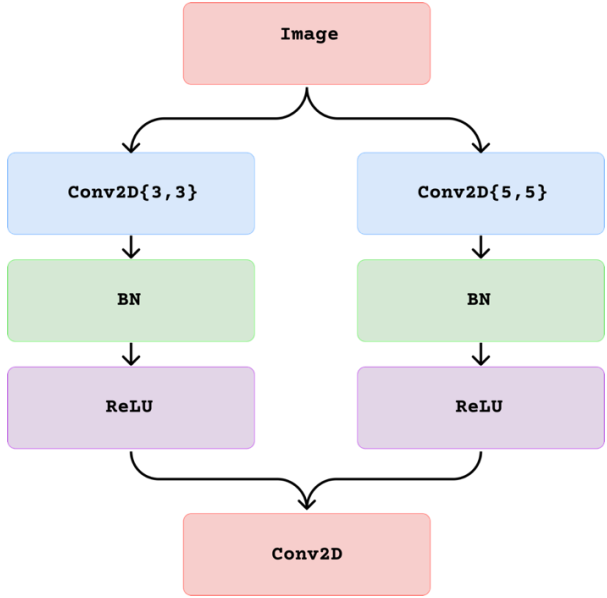

**Figure 1**: Architecture design of Dual Pathway/Head convolution and its concatenation.

The architecture is expressed mathematically as:

$$h(X) = G_1(H_1(X;\theta_1);\theta_2) + G_2(H_2(X;\theta_3);\theta_4) \quad (1)$$

$$F(X) = G_3(H_3(h(X);\theta_5);\theta_6) \quad (2)$$

where $F$ denotes the DPN, $X$ denotes the input signal, $H_1$ and $H_2$ represent the distinct pathways, $G_1$ and $G_2$ denote the transformation functions, and $\theta$ represents the parameters of the respective layers. The primary pathway, $H_1$, is architected to extract fine-grained, localized features, while the auxiliary pathway, $H_2$, is geared towards encapsulating broader, global features. A more complete representation of features is possible due to the dual-pathway design.

### 3.3. Self-Attentive Canonical Correlation (SACC)

#### 3.3.1. CCA & Self Attention Theoretical Background

*Canonical Correlation Analysis (CCA)* [31] is a well-known statistical approach for finding the linear relationship between two multivariate datasets. Let $X \in R^{n\times p}$ and $Y \in R^{n\times q}$ are two matrices containing two sets of variables with $n$ observations. CCA seeks linear combinations of the variables in $X$ and $Y$ that have maximum correlation with each other. Let, $x$ be a linear combination of columns of $X$ and $y$ a linear combination of columns of $Y$:

$$x = X\boldsymbol{a}, \quad y = Y\boldsymbol{b} \quad (3)$$

where $a \in R^p$ and $b \in R^q$ are vectors of coefficients.

CCA attempts to find the vectors $a$ and $b$ that maximize the correlation between $x$ and $y$:

$$\rho = \frac{\text{cov}(x,y)}{\sqrt{\text{var}(x)\cdot \text{var}(y)}} \quad (4)$$

*Self-Attention* [32] is expressed as follows: Given a sequence of input vectors $\{x_1, x_2, \ldots, x_n\}$, the self-attention mechanism computes a new sequence of vectors $\{y_1, y_2, \ldots, y_n\}$ where each vector $y_i$ is a linear combination of all the input vectors $x_i$, weighted by attention scores. These attention scores are derived from a compatibility function of the input vectors, often parameterized by learnable weights. The following equation governs the computation of the attention scores:

$$\alpha_{ij} = \frac{\exp(\text{score}(x_i, x_j))}{\sum_k \exp(\text{score}(x_i, x_k))} \quad (5)$$

Here, $\text{score}(x_i, x_j)$ is a function that computes the compatibility between $x_i$ and $x_j$, commonly implemented using a dot product or a feed-forward neural network. Following the computation of the attention scores, the output vectors $y_i$ are computed as a weighted sum of the input vectors $x_j$, with the weights being the attention scores $\alpha_{ij}$:

$$y_i = \sum_j \alpha_{ij} \cdot x_j \quad (6)$$

### 3.3.2. SACC Architecture implementation

As stated earlier, various medical studies suggest the influence of patient characteristics ($P_C$) on heart conditions. For example, smoking habits degrade the heart's capacity to pump blood [33]. So, it is essential to study the effect of $P_C$ on ECG morphology ($E_M$) for a more personalized diagnosis. To bridge the gap in the correlation analysis of $P_C$ and $E_M$, we introduced the SACC (Self-Attentive Canonical Correlation) layer to learn $P_C$'s influence on $E_M$ explicitly.

**Table 1**
Pseudocode for Self-Attentive Canonical Correlation layer of the rECGnition_v2.0 algorithm.

| **Algorithm**: $Self-Attentive\ Canonical\ Correlation$ (SACC) |
|---|
| **Input**: $E_m$ feature map $X$, $P_c$ feature map $Y$<br>**Output**: Canonically correlated vectors $XY$ |
| 1: Initialize canonical weight matrix a and b with random values |
| 2: Repeat until convergence: |
| 3:     Update a and b by solving: |
| 4:         $\max_{a,b} \text{corr}(aX, bY)$ |
| 5:     Apply self-attention on $aX$ and $bY$ to obtain $X'$ and $Y'$ |
| 6:     Apply non linearity on $X'$ and $Y'$ to get $XY$ |
| 7: End Repeat |
| 8: Return $XY$ |

This approach combines the robustness of Canonical Correlation Analysis (CCA) with the attention mechanism characteristics of self-attention resulting in a better feature fusion paradigm. The CCA framework as explained earlier, was integrated with the self-attention mechanism (to better correlate between $P_C$ and $E_M$) that gave rise to novel deep learning layer termed as SACC. The SACC layer can be articulated through the pseudocode (Table 1), and its neural network architectural design (Figure 2). Moreover, it was implemented using custom TensorFlow-Keras layer. SACC projects two input tensors; representing ECG signal features and patient metadata features into a shared latent space, applies a self-attention mechanism to weigh the importance of each feature, and then fuses these representations through a non-linear transformation. Mathematical operations that are being sequentially performed inside the layer are as follows: Let $F_{img} \in R^{1 \times d_1}$ and $F_{pat} \in R^{1 \times d_2}$ represents the ECG signal features and patient metadata features, respectively, where $d_1, d_2$ are the feature dimensions. Initially, the SACC layer linearly projects these inputs into a common latent space of dimension N (where N is a non-trainable parameter given during model initialization), using weight matrices $W_{img} \in R^{d_1 \times N}$ and $W_{pat} \in R^{d_2 \times N}$. Next to add non-linearity, the $ReLU$ activation was employed on the projections as $Z_{img} = \text{ReLU}(F_{img} W_{img})$ and $Z_{pat} = \text{ReLU}(F_{pat} W_{pat})$. Furthermore, attention scores were computed using SoftMax normalization of learned attention weights as $\alpha_{img} \in R^N$ and $\alpha_{pat} \in R^N$; $\text{Attention}_{img} = \text{softmax}(\alpha_{img})$ and $\text{Attention}_{pat} = \text{softmax}(\alpha_{pat})$. These attention scores modulate the projected representations as $Z_{pat} = \text{Attention}_{img} \odot Z_{img}$ and $Z_{pat} = \text{Attention}_{pat} \odot Z_{pat}$, where $\odot$ denotes element-wise multiplication. The modulated representations were then concatenated and passed through a dense layer with a ReLU activation function to perform non-linear fusion; $F = \text{ReLU}(W_f[Z_{img}; Z_{pat}])$, where $W_f$ represents the weights of the dense layer. The output $F$ of the SACC layer has the shape $(1, N)$, effectively combining the information from both input modalities into a coherent and integrated representation.

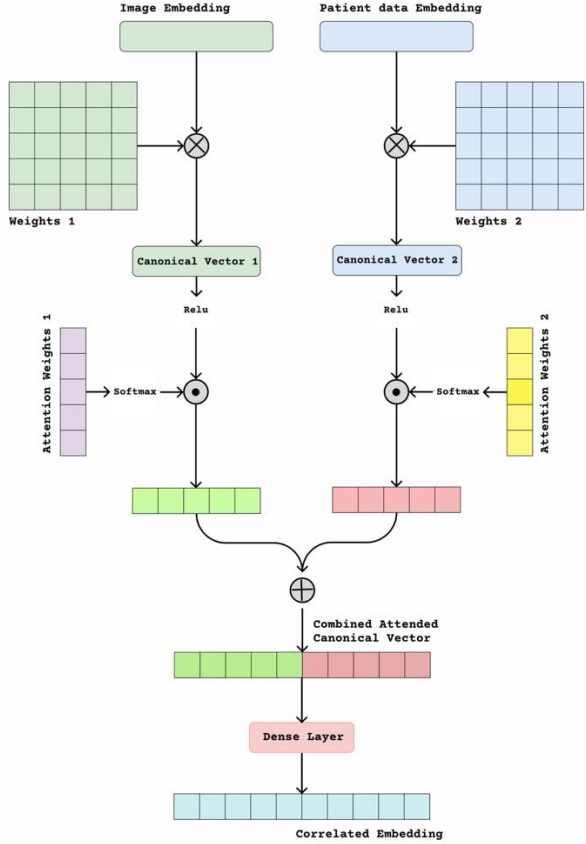

**Figure 2**: Architecture design of Self-Attentive Canonical Correlation layer (SACC).

### 3.4. rECGnition_v2.0 Architecture implementation

A novel model architecture (Figure 3, Table 2) that combines depth-wise separable convolution, DPN, and SACC was developed.

**Table 2**
rECGnition_v2.0 body architecture.

| Type / Stride | Filter Shape | Output Shape |
|---|---|---|
| Input Image | - | 128 x 128 x 1 |
| Conv 3x3 / s2 | 3 x 3 x 1 x 32 | 64 x 64 x 32 |
| Conv 5x5 / s2 | 5 x 5 x 1 x 32 | 64 x 64 x 32 |
| Concat | - | 64 x 64 x 64 |
| Dw-Conv / s1 | 3 x 3 x 64 dw | 64 x 64 x 64 |
| Dw-Conv / s2 (x2) | 3 x 3 x 128 dw | 16 x 16 x 128 |
| Dw-Conv / s2 (x3) | 3 x 3 x 256 dw | 2 x 2 x 256 |
| MaxPool | Pool 2x2 | 1 x 256 |
| Input $P_E$ | - | 1 x 2 |
| Dense $P_E$ | 2x8 | 1 x 8 |
| Concat | - | 1 x (10) |
| SACC Layer | 256x256 (x3) | 1 x 256 |
| FC Layer 1 | 256x128 | 1 x 128 |
| FC Layer 2 | 128x64 | 1 x 64 |
| SoftMax Classifier | 64x10 | 1x10 |

*s1*: (1, 1) stride; *s2*: (2, 2) stride; Dw-Conv: Depth-wise separable convolution

This model, herein referred to as the rECGnition_v2.0 (robust ECG abnormality detection) algorithm, conducts a balanced convergence of feature fusion, pathway-based analysis, and computationally efficient convolution operations, establishing a robust framework for insightful cardiac signal analysis. The architecture of rECGnition_v2.0 was carefully designed to make the most of the strengths of each of its constituent methodologies. We have developed the two variants for rECGnition_v2.0. The smaller variant is used for benchmarking and comparative analysis, while the medium variant is designed to demonstrate the model's scalability. The medium variant uses bigger input size with 512 features, on the other hand smaller variant uses 256 features. The first neck i.e. ECG heartbeat processor (Figure 3A) processes the input image via two parallel ConvB (Figure 3B) layers with 3x3 and 5x5 kernel sizes, respectively, each yielding 32 feature maps (Table 2). These maps were concatenated to form a unified feature space, which was then funnelled through a series of depth-wise separable convolutional i.e. Dw-ConvB (Figure 3C) blocks with an ascending filter count: 64, 128, to 256. Dropout layers were intermittently employed to mitigate overfitting at two places.

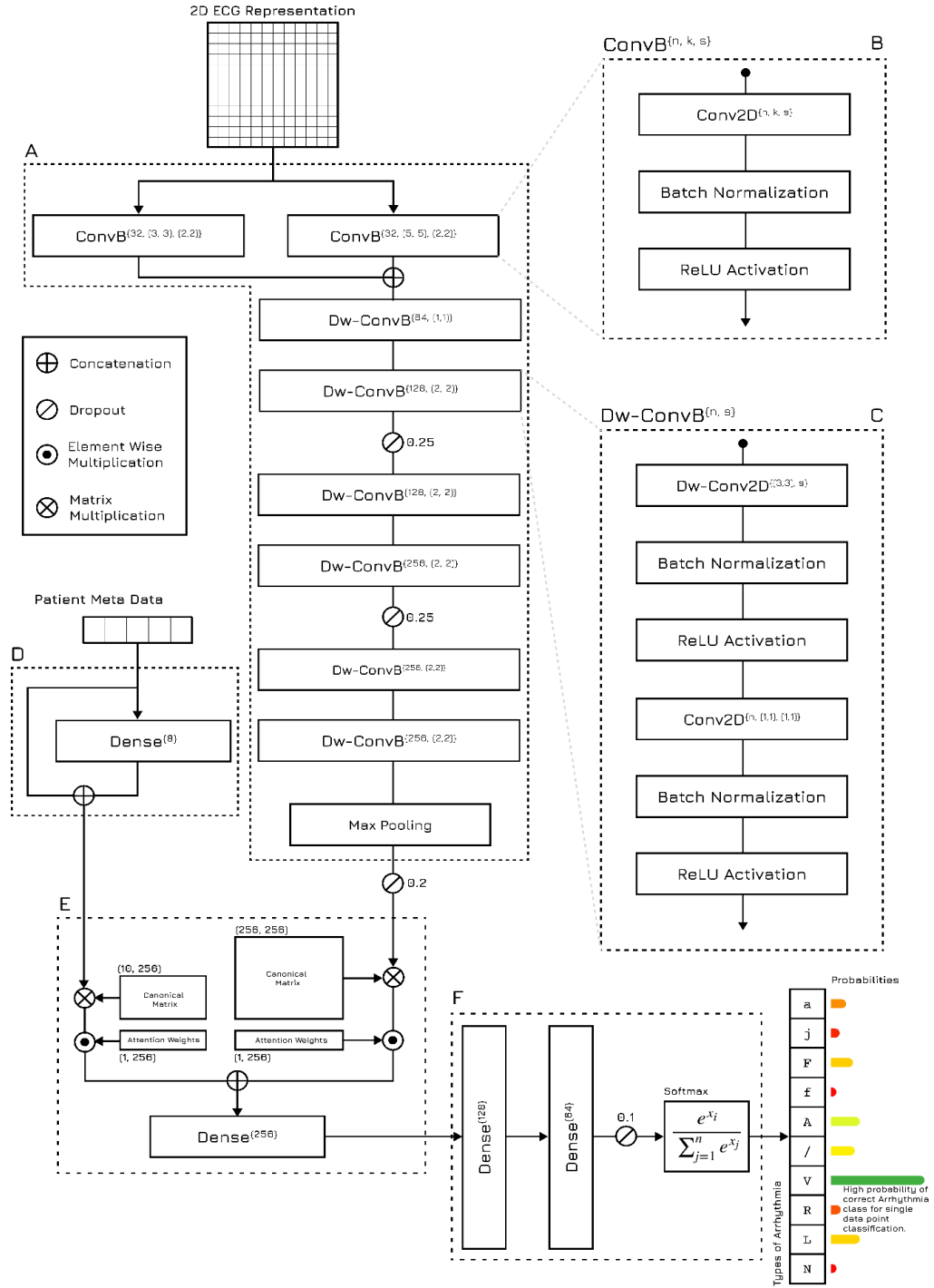

**Figure 3**: Structural diagram of proposed rECGnition_v2.0 model. (A) represents the ECG heartbeat processor unit which was implemented using ConvB and Dw-ConvB modules, (B) represents the internals of the ConvB modules and (C) represents the internals of Dw-ConvB module where *n* is number of filters, *k* is kernel size, and *s* is stride size, (D) represents the meta data processor unit. The output from (A) and (D) is fed into SACC layer represented by (E). Finally, the output of SACC is forwarded to (F) which is the meta classifier responsible for predicting the probabilities of corresponding arrhythmia classes.

The second neck i.e. meta data processor (Figure 3D), dedicated to patient characteristics, utilizes a single dense layer with 8 neurons to generate a feature map, which was then fused back with the input to produce the final patient feature representation. The output feature maps from both paths, lets denote them as $F_{img}$ and $F_{pat}$, were given as input into the SACC layer (Figure 3E; in-depth implementation details are mentioned in previous paragraph) by ensuring the dimensionality of canonical matrices $HXW$ satisfies $W \geq \max\left(\dim(F_{img}), \dim(F_{pat})\right)$ to preserve feature integrity. The SACC layer outputs a 256-dimensional canonical vector, which is projected through a neural network with dense layers of 128 and 64 neurons, closes with a SoftMax-activated output layer that assigns probabilities to the respective arrhythmia classes (Figure 3F). The model thus elegantly merges convolutional feature extraction with patient data analysis, using the SACC layer to perform attention-augmented canonical correlation for superior prediction performance. The abstract model representation of the model is illustrated in Figure 4.

### 3.4.1. rECGnition_v2.0 Computational Efficiency Considerations

The theoretical computational complexity assessment was required as this study tried to create a computationally efficient model for arrhythmia classification. Hence, we have extensively analysed the constituting rECGnition_v2.0 components to find out the Big-O complexity in terms of input image size and kernel size. The proposed architecture can be broken down into 3 major blocks, each distinguished by its unique operational methodology. So, we analysed each blocks computational complexity separately.

***Initial convolution used with different receptive fields (DPN)*** have the complexity of a simple convolution operation performed on an input size of $H' \cdot W'$ with $K_1$ and $K_2$ kernel sizes. Each element in the output feature map is obtained by performing a convolution operation which entails $K^2 \cdot N$ multiplications and additions where $N$ is the number of channels. Given that each filter traverses the entire input matrix to generate the corresponding output feature map, the total

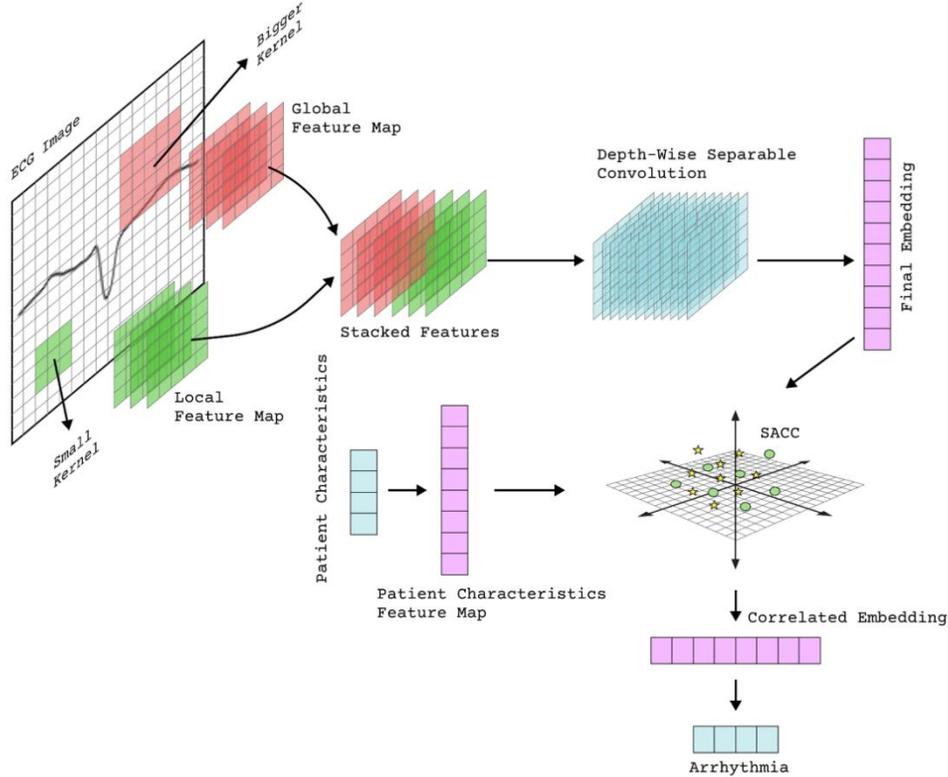

**Figure 4**: Graphical representation of rECGnition_v2.0's algorithm.

number of operations for all filters $F$ is $O(N \cdot (K_1^2 + K_2^2) \cdot F \cdot H' \cdot W')$.

***Intermediate layers, which use depth-wise separable convolution*** (a variant of the standard convolution operation) comprises of two distinct steps: a depth-wise convolution followed by a pointwise convolution.

**Depth-wise Convolution:** The depth-wise convolution operates on each input channel separately with a distinct kernel per channel. The computational complexity for this phase can be encapsulated as $O(N \cdot K^2 \cdot H' \cdot W')$.

**Pointwise Convolution:** The pointwise convolution is essentially a $(1 \times 1)$ convolution that combines the outputs of the depthwise convolution across channels. The computational complexity for this phase is given by $O(N \cdot M \cdot H' \cdot W')$.

The aggregate computational complexity of the depth-wise separable convolution is the sum of the complexities of the depth-wise and pointwise convolutions, which is $O(N \cdot K^2 \cdot H' \cdot W' + N \cdot M \cdot H' \cdot W')$.

***The SACC layer*** is the combination of attended feature projections merged using a fully connected layer. Thus, the computational complexity for taking projection operations is $O(D_1^2 \cdot M + D_2^2 \cdot M)$ and applying attention map to it is $O(2M)$. The dense layer operation is also a matrix multiplication with a complexity of $O(2M \cdot M)$ where $D_1$ and $D_2$ are the dimension of input vectors to the SACC, and $M$ represents the output dimension of the SACC.

### 3.5. Data Acquisition and Pre-processing

This study exploited the MITDB [34, 35] for experimentation and model evaluation.

INCARTDB and EDB datasets were further used to strengthen the generalizability and transferability of rECGnition_v2.0. Leveraging the R-peak annotations provided within the dataset, each cardiac signal was segmented into individual heartbeats. The segmentation was executed as per the following equation:

$$x(n) = \{ x[n] \mid n = R\_peak - \Delta n \text{ to } R\_peak + \Delta n \} \quad (7)$$

where $x(n)$ delineates the extracted heartbeat, $R_{peak}$ signifies the R-peak location and $\Delta n$ demarcates a predefined window around the R-peak. A crucial step was employed to lessen the effects of inherent noise within the data using convolution-based denoising techniques along with Hann window [36], renowned for its efficacy in attenuating signal noise while preserving the essential characteristics of the underlying signal.

The following mathematical expression represents the convolution operation employed for denoising:

$$y[n] = \sum_{k=-\infty}^{\infty} x[k] \cdot w[n-k] \quad (8)$$

where $y[n]$ denotes the denoised signal, $x[k]$ represents the original signal, and $w[n-k]$ is the Hann window defined as:

$$w[n] = 0.5 \cdot (1 - \cos(2\pi n/N)) \text{ for } 0 \leq n \leq N \quad (9)$$

The Hann window was chosen for this denoising application because its smooth, cosine-shaped tapering reduces spectral leakage [37]. This operation is particularly effective in getting rid of loud transient noises, thereby refining the signal quality for subsequent analysis. Comparing cardiac signals pre- and post-denoising showed the efficacy of convolution-based denoising leveraging the Hann window. This comparison (Suppl. Figure S1) clearly demonstrated noise attenuation and the preservation of the fundamental signal characteristics essential for accurate arrhythmia classification.

## 4. Experiment Setup

In our experimental setup, we trained rECGnition_v2.0 with the following hyperparameters on Google colab GPUs and TPUs (TPUs were primarily used in early development of the model development due to TPU's faster training time): a batch size of 32, training for 40 epochs (1570 steps per epoch), and an image input size of 128x128 pixels. The initial learning rate was set to 0.01, and the Adam optimizer (described in detail in section 4.1) was employed for parameter updates. To ensure reproducibility, the random seed was fixed at 257. To enhance the learning process, we integrated a cosine learning rate scheduler with warmup steps. Specifically, the learning rate scheduler was configured with an initial learning rate of 0.01 and warmup steps was 5. The scheduler follows a cosine decay pattern with 0.5 cycles over the course of the training. The learning rate $\eta(t)$ at epoch $t$ was computed as follows:

$$\eta(t) = \begin{cases} \frac{t}{\max(5, \text{num\_warmup\_steps})} \times \text{lr} & \text{if } t < \text{num\_warmup\_steps} \\ \text{lr} \times 0.5 \left(1 + \cos\left(\frac{\pi \times \text{num\_cycles} \times 2 \times (t - \text{num\_warmup\_steps})}{\text{num\_training\_steps} - \text{num\_warmup\_steps}}\right)\right) & \text{otherwise} \end{cases}$$

where num_training_steps = EPOCHES. This learning rate schedule allows for a smooth adjustment of the learning rate, starting with a gradual warmup phase followed by cosine decay, which helps in stabilizing the training process and potentially improving model performance. The training convergence of categorical cross-entropy loss is given in Suppl. Figure S2. Moreover, the total dataset was splitted into 9:1 ratio of training and testing.

## 4.1 Optimization

Optimization is pivotal in navigating the vast trainable parameter space to find a set of trainable parameters that minimize the loss function, thereby ensuring the model's adeptness in making accurate predictions. The Adam optimizer, known for its efficacy in both sparse gradients and noisy data scenarios, was employed in this study. The Adam optimizer operates as per the following update rules:

$$m_t = \beta_1 m_{t-1} + (1 - \beta_1) g_t \quad (11)$$

$$v_t = \beta_2 v_{t-1} + (1 - \beta_2) g_t^2 \quad (12)$$

$$\widehat{v_t} = \frac{v_t}{1 - \beta_2^t} \quad (13)$$

$$\widehat{m_t} = \frac{m_t}{1 - \beta_1^t} \quad (14)$$

$$\theta_t = \theta_{t-1} - \alpha \frac{\widehat{m_t}}{\sqrt{\widehat{v_t}} + \epsilon} \quad (15)$$

Where $\theta$ represents the parameters, $g$ represents the gradients, $m$ and $v$ are moment estimates, $\beta_1, \beta_2$ are the exponential decay rates for the moment estimates, $\alpha$ is the learning rate and $\epsilon$ is a small value to prevent division by zero.

## 4.2 Regularization

Regularization is standard for preventing overfitting, especially in scenarios where the model's complexity is high. Dropout is a basic yet effective regularization technique that randomly eliminates a portion of the neurons during training, preventing any single neuron from becoming overly specialized. The dropout operation is expressed as:

$$y_i = x_i \cdot z_i \quad (16)$$

where $x_i$ is the input to a neuron, $y_i$ is the output, and $z_i$ is a Bernoulli random variable with probability $p$ of being 1. Suppl. Table S1 summarises the ablation study for different dropout values used inside the image processor unit and near the output layer to study the effect of regularization on rECGnition_v2.0's performance.

## 5. Results and Discussion
### 5.1. Model Performance Evaluation

To assess the efficacy of rECGnition_v2.0, the precision (P), recall (R), and F1-score were evaluated for each class. For the class '*f*', the rECGnition_v2.0 achieved perfect precision, recall, and F1-score (Table 3), suggesting that it is exceptionally proficient in identifying this category. A similar trend of high performance was observed in classes '*L*', '*N*', '*R*', and '*V*', demonstrating precision and recall above 95%, indicative of the robustness of the rECGnition_v2.0 framework. The medium variant of the model, also showed similar/high precision and F1-score, demonstrating scalability of the model (Suppl. Table S3). Figure 5 represents the ROC-AUC curve of

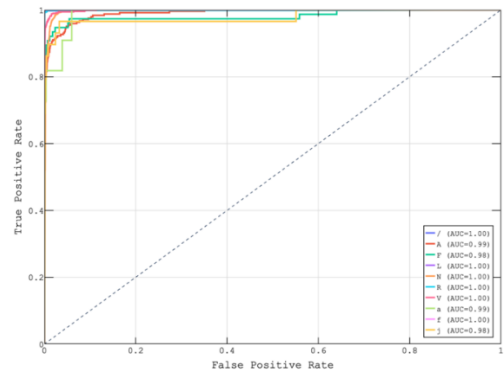

**Figure 5**: ROC-AUC plot for 10 arrhythmia classification using rECGnition_v2.0. *N*: Normal beat; *L*: Left Bundle Branch Block beat; *R*: Right Bundle Branch Block beat; *V*: Premature Ventricular Contraction beat; /: Paced beat; *A*: Atrial Premature beat; *f*: Fusion of paced and normal beat; *F*: Fusion of ventricular and normal beat; *j*: Nodal escape beat; *a*: Aberrated atrial premature beat.

arrhythmia classification of the model with an almost perfect score of 1.0 for 6 out of 10 classes and ≥ 0.98 scores for all types of arrhythmias. In our experiments, we evaluated the performance of three type of fusions Simple Concatenation, CCA and SACC across precision, recall, and F1-score metrics which revealed that SACC exhibits superior performance metrics across nearly all arrhythmia classes (Table 3). The paired *t*-test results demonstrated that the SACC model significantly outperforms the CCA model in terms of F1-score ($p = 0.0386$). Also, the SACC model exhibited a trend towards higher F1-

**Table 3**
Comparison of performance results for various fusion techniques for 10 class arrhythmia classification.

| Class | Simple Concatenation | | | CCA | | | SACC (rECGnition_v2.0) | | |
|---|---|---|---|---|---|---|---|---|---|
| | P | R | F1 | P | R | F1 | P | R | F1 |
| / | 99.48 | 99.74 | 99.61 | **100.0** | 99.48 | 99.74 | 99.74 | **99.74** | **99.74** |
| A | 91.89 | 83.27 | 87.37 | **96.08** | 60.0 | 73.87 | 92.54 | **86.12** | **89.22** |
| F | 85.51 | 77.63 | 81.38 | 90.48 | 75.0 | 82.01 | **91.04** | **80.26** | **85.31** |
| L | 99.75 | 99.88 | 99.81 | 99.38 | 99.75 | 99.56 | **99.75** | **99.88** | **99.81** |
| N | 97.46 | 98.95 | 98.20 | 95.95 | 99.17 | 97.54 | **98.32** | **99.05** | **98.68** |
| R | **99.66** | 98.63 | 99.14 | 98.81 | 98.98 | 98.89 | 99.32 | **99.15** | **99.23** |
| V | 95.47 | 94.67 | 95.07 | 93.53 | 95.23 | 94.37 | **96.78** | **96.91** | **96.85** |
| a | 66.67 | 54.55 | 60.0 | **100.0** | 36.36 | 53.33 | 66.67 | **90.91** | **76.92** |
| f | 95.24 | 95.24 | 95.24 | 87.50 | 100.0 | 93.33 | **100.0** | **100.0** | **100.0** |
| j | **90.48** | 65.52 | 76.00 | 71.43 | 34.48 | 46.51 | 80.77 | **72.41** | **76.36** |

*N*: Normal beat; *L*: Left Bundle Branch Block beat; *R*: Right Bundle Branch Block beat; *V*: Premature Ventricular Contraction beat; /: Paced beat; *A*: Atrial Premature beat; *f*: Fusion of paced and normal beat; *F*: Fusion of ventricular and normal beat; *j*: Nodal escape beat; *a*: Aberrated atrial premature beat.

scores compared to Simple Concatenation (Table 3). This denotes that the integration of features through SACC not only captures the underlying association between the $P_C$ and $E_M$ more effectively but also enhances the discriminative power of the model.

### 5.2. Enhanced Feature Representation and Importance

SACC is a better way to figure out how patient-specific factors like age and gender affect the shape of an ECG heartbeat. The self-attention mechanism in SACC provides a glimpse into the feature importance. Figure 6 represents the activation regions of the heartbeat for only image feature embedding and embedding generated after applying SACC; it shows that the model has learned to select the critical regions like the start of different $E_M$ segments, location of changes, etc. (which define the structure of the PQRST construct) more precisely.

The PCA[rECGnition_v2.0] of distinct classes of arrhythmia peeks into the inner working of rECGnition_v2.0 which exhibited a degree of linear separability, particularly for certain classes such as '*f*' and '*j*', which align along unique trajectories (Figure 7). This indicates that the rECGnition_v2.0 model has learned to extract features that result in a feature space where arrhythmia classes are distinctly separable, potentially facilitating higher classification performance for these classes; for instance, 100% prediction score for '*f*' (Table 3). This unique behaviour of induced linear separability by virtue of SACC helped rECGnition_v2.0 in yielding state-of-the-art accuracy by keeping the model parameters lowest. Conversely, the PCA [EfficientNetB0] presents a more entangled representation, with significant overlap between classes.

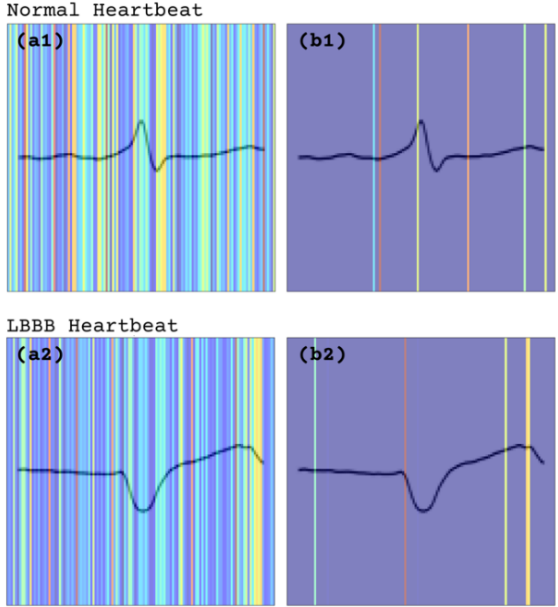

**Figure 6**: Comparison of activation locations for the ECG image before and after SACC; (a1, a2) represents activation for Normal and LBBB heartbeat respectively before SACC, (b1, b2) represents activation for Normal and LBBB respectively, after SACC.

Despite this, certain clusters can be seen within the data, suggesting that while the classes are not as linearly separable as in the rECGnition_v2.0 transformed space, there still exists a structure that could be exploited by a classifier, but the model needs to learn more complex boundaries to differentiate between classes effectively.

### 5.3. Additional Experiments

Various set of additional experiments were performed to better understand the rECGnition_v2.0 results, *i.e.* compared the performance of with and without meta data, and carried out the ablation study of DPN. To further strengthen the performance and generalization capabilities of rECGnition_v2.0 we expanded our experiment to include the INCARTDB and EDB dataset. Additionally, we also compared rECGnition_v2.0 with SOTA models which showed rECGnition_v2.0 exhibits the improved efficacy for these set of experiments as well.

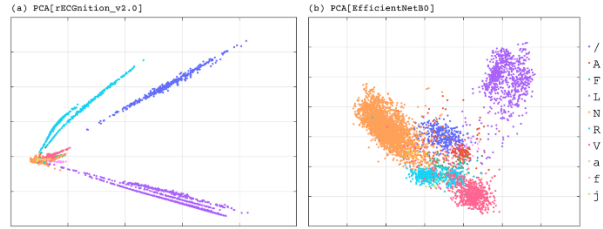

**Figure 7**: Principal component analysis of final embeddings generated by rECGnition_v2.0 and EfficientNetB0. *N*: Normal beat; *L*: Left Bundle Branch Block beat; *R*: Right Bundle Branch Block beat; *V*: Premature Ventricular Contraction beat; */*: Paced beat; *A*: Atrial Premature beat; *f*: Fusion of paced and normal beat; *F*: Fusion of ventricular and normal beat; *j*: Nodal escape beat; *a*: Aberrated atrial premature beat.

#### 5.3.1. With and without patient data analysis

For arrhythmia detection, two conditions were compared: one considering only ECG data and another incorporating patient characteristics. The plot (Suppl. Figure S3) reveals that the integration of patient characteristics generally enhances the F1-score across most arrhythmia classes supporting our methodology, with notable improvements in classes '*A*', '*F*', '*V*', '*f*', and '*j*'. This enhancement underscores the value of including demographic parameters along with ECG in deep learning models. Particularly, for class '*F*', there is a significant leap in F1-score when patient characteristics are included, suggesting that additional patient data provides critical context for more accurate arrhythmia classification.

### 5.3.2. Analysis of the Dual Pathway Mechanism

The deeper analysis of dual pathway mechanism was performed to test influence of shorter and longer receptive fields.

When compared with single pathway *i.e.* only smaller receptive fields, DPN demonstrated improved performance; the improvement is apparent where single pathway poorly performs such as '*a*' and '*j*' classes (Suppl. Table S4). Using Grad-Cam visualization, we observed that bigger receptive fields tries to discard local changes (Figure 8), the bigger kernel's individual feature map has more smoothened and dilated ECG signal construct which shows aggregation of information at larger scale, by discarding local aberrations.

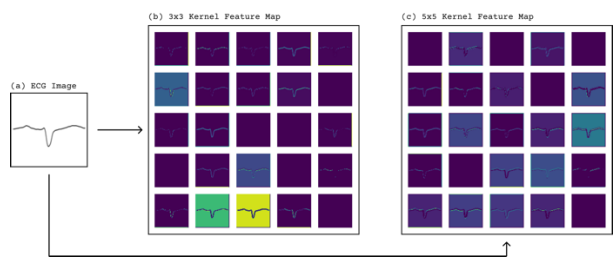

**Figure 8**: Feature map visualization after applying different receptive fields, (a) segmented ECG heartbeat, (b) 25 features out of 32 features extracted using 3x3 kernel, (c) 25 features out of 32 features extracted using 5x5 kernel.

### 5.3.3. Comparative analysis with other SOTA CNN models

The rECGnition_v2.0 models were compared against *MobileNet*, *ResNet50* and *EfficientNetB0* to determine its performance characteristics. The accuracy is highest among all for rECGnition_v2.0 (Suppl. Table S5) with the lowest FLOPs, trainable parameters, and maximum predictions per second depicting rECGnition_v2.0's ability to outperform SOTA CNN models for ECG classification task with lowest computational requirement and maximum speed. While overall accuracy increased only by 1% for the proposed model, which might not seem so significant, the model has performed remarkably well for arrhythmia classes that have lower occurrences (Figure 9, Table 3).

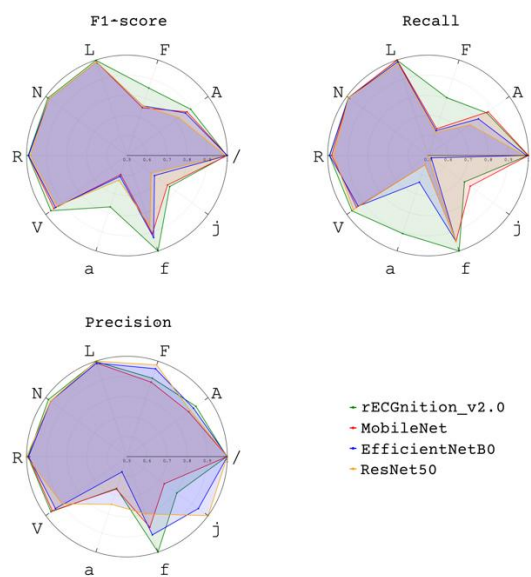

**Figure 9**: Comparison of Precision, Recall, and F1-score for 10 Arrhythmia classification using MobileNet, EfficientNetB0, ResNet50, and rECGnition_v2.0.

### 5.3.4. Computational Efficiency Analysis

For real-time clinical implementation of such a model, low computational requirement of the model is an essential need. rECGnition_v2.0 is designed and developed with the aim of clinical deployment; thus, it obeys the low computational requirement for the model. As compared with other SOTA CNN-based models, we can see that rECGnition_v2.0 is ahead of those models by a significant margin in terms of memory and low parameter/FLOPs (Suppl. Figure S4, Table S5).

*5.3.5. Generalization Assessment*

We have performed additional set of experiments on INCARTDB [35] and EDB [38], which are different datasets, and on MITDB on AAMI beats. This assessment setup aimed to showcase rECGnition_v2.0's ability to generalize on wide range of data distributions. As shown in Table 4, the empirical evidence suggests that the model not only preserves its predictive accuracy but also adapts to the intricacies and variations presented by different classification datasets/groups. Furthermore, it concluded that the model didn't overfit the MITDB dataset.

*5.4 Comparative analysis*

A thorough comparison of the rECGnition_v2.0 model with other advanced approaches for arrhythmia classification was carried out (Table 5). The benchmark datasets used for evaluation were MITDB, INCARTDB and EDB. The basis of our comparison revolved on many crucial aspects, including the total F1-score, prediction speed, computing efficiency, and generalization on different dataset distributions. The rECGnition_v2.0 model exhibited outstanding performance in terms of F1-score, attaining remarkable values of 99.01%, 98.21%, and 98.05% in the 5-class, 8-class, and 10-class arrhythmia classification tasks, respectively. These findings highlight the model's exceptional accuracy and ability to perform well across different degrees of categorization complexity. One notable feature of rECGnition_v2.0 is its exceptional prediction speed, averaging around 34 milliseconds per input sample. This improved efficiency is a notable progress in enabling real-time diagnosis of arrhythmia, which has the potential to improve patient outcomes by allowing prompt intervention. Houssein *et al*. [39] and Chen G *et al*. [40] proposed work has depicted better F1-scores than rECGnition_v2.0 because of the hand-crafted features used in both the methodology; although the overall prediction capability of model is more but the hand-crafted feature methodologies are more prone to errors as any failure in feature extraction will propagate to classifier [41]. The LSTM-Autoencoder model proposed by Liu *et al*. [42] has shown reduced computational requirements and achieved an F1-score of 97.75%. However, it is crucial to acknowledge that its sequential data processing leads to prediction speeds that are five times slower compared to rECGnition_v2.0.

**Table 4**
Generalization results of rECGnition_v2.0 for classifying as per AAMI standards on INCARTDB, MITDB and EDB datasets.

| Class | INCARTDB | | | MITDB | | | EDB | | |
| --- | --- | --- | --- | --- | --- | --- | --- | --- | --- |
| | *P* | *R* | *F1* | *P* | *R* | *F1* | *P* | *R* | *F1* |
| **Normal** | 99.39 | 99.63 | 99.51 | 99.17 | 99.38 | 99.27 | 96.93 | 98.93 | 97.92 |
| **SEB** | 96.20 | 89.39 | 92.67 | 91.26 | 90.94 | 91.10 | 84.38 | 58.70 | 69.23 |
| **VEB** | 99.55 | 99.75 | 99.65 | 97.76 | 97.62 | 97.69 | 97.89 | 97.61 | 97.75 |

*Normal*: Normal beat; *SEB:* Supraventricular Ectopic Beat*; VEB:* Ventricular Ectopic Beat

**Table 5**

Comparative study of rECGnition_v2.0 with the previously developed models on 5/8/10 class non-AAMI and AAMI arrhythmia classification.

| | Database | Publications | CI[β] | Classifier | Classes | Model Characteristics[##] | | | Performance | |
|---|---|---|---|---|---|---|---|---|---|---|
| | | | | | | No. of Params | Flops | Time[♩] | Acc | F1 |
| Non-AAMI | MITDB | Melgani et al. [43] | 24.4 | SVM | 6 | - | - | - | 91.67 | - |
| | | Dutta et al. [44] | 9.6 | least square support vector machine | 3 | - | - | - | 95.82 | 91.00 |
| | | Jun et al. [45] | 49.2 | 2D CNN – VGGNet | 8 | 14.7M | 39.9b | 395ms | 98.85 | 97.12 |
| | | Midani. Et al. [46] | 14.4 | Sequential fusion method to combine feed-forward and recurrent deep neural networks | 5 | 2.1M | 4.9G | 73ms | 99.46 | 97.63 |
| | | Shofiqul et al. [47] | 5.4 | Hierarchical attention based dual structured RNN with dilated CNN | 5 | 2.8M | 50.2B | 365ms | 99.01 | 97.51 |
| | | Mathunjwa et al. [48] | 44.0 | ResNet-18 | 4 | 11.4M | 1.9B | 95ms | 97.21 | 95.96 |
| | | Liu et al. [42] | 5.1 | LSTM, Autoencoder | 5 | 21.2M | 42.2M | 165ms | 98.57 | 97.75 |
| | | Amin et al. [49] | 48.2 | Vanilla CNN | 8 | 10.1M | 7.2b | 140ms | 99.11 | 98.00 |
| | | Amin et al. [49] | 48.2 | Vgg16 | 8 | 14.7M | 39.9b | 395ms | 98.90 | 97.00 |
| | | Zhang, et al. [50] | 7 | Time–frequency features inputted to the CNN | 8 | - | - | - | 99.10 | - |
| | | Jha et al. [51] | 20.4 | TQWT-based features of ECG beats and SVM classifier | 8 | - | - | - | 99.27 | 96.22 |
| | | **Proposed** | - | Self-attentive canonical fusion & CNN | 5 | **450k** | **82.7M** | **34ms** | **99.25** | **99.01** |
| | | **Proposed** | - | Self-attentive canonical fusion & CNN | 8 | **450k** | **82.7M** | **34ms** | **98.26** | **98.21** |
| | | **Proposed** | - | Self-attentive canonical fusion & CNN | 10 | **450k** | **82.7M** | **34ms** | **98.07** | **98.05** |
| AAMI | MITDB | Ince et al. [52] | 137.4 | 1-D CNN | 5 | - | - | - | 96.40 | 73.02 |
| | | Ortín, S., et al. [53] | 5.1 | Echo State Network (ESN) | 2 | - | - | - | - | 95.20 |
| | | w, z., et al. [54] | 13.5 | ECG Morphology and Segment Feature with SVM-RBF | 4 | - | - | - | 97.80 | - |
| | | Cui et al. [55] | 12.4 | 1-D CNN + DWT + PCA | 5 | - | - | - | 98.35 | 98.36 |
| | | Li, Zhou et al. [56] | 25.4 | End-to-end 31-layer deep *ResNet* | 4 | 25M | 3.6B | 147ms | 99.06 | 97.30 |
| | | Sinha, N. et al. [57] | 7.2 | MSC, MTF & MPC using DNN classifier | 5 | - | - | - | 99.05 | - |
| | | Houssein et al. [39] | 27.7 | IMPA-CNN | 4 | - | - | - | 99.33 | 98.65[#] |
| | | **Proposed** | - | Self-attentive canonical fusion & CNN | 4 | **450k** | **82.7M** | **34ms** | **98.60** | **98.60**[*] |
| | INCARTDB | Liu Y et al. [58] | 5.1 | CNN + Per patient training | 5 | 14.7M | 39.9b | 395ms | 97.11 | 96.57 |
| | | Houssein et al. [39] | 27.7 | IMPA-CNN | 4 | - | - | - | 99.43 | 98.86[#] |
| | | Wang, G et al. [25] | 16.3 | Domain adoption | 2 | - | - | - | 95.36 | - |
| | | V. Kalidas et al. [59] | 1.4 | AE + RF | 2 | - | - | - | - | 91.30 |
| | | Chen G et al. [40] | 5.7 | Feature fusion, cascaded classifier | 4 | - | - | - | - | 99.80[#] |
| | | **Proposed** | - | Self-attentive canonical fusion & CNN | 4 | **450k** | **82.7M** | **34ms** | **98.68** | **98.01**[**] |
| | EDB | Houssein et al. [39, 60] | 27.7 | IMPA-CNN | 4 | **-** | **-** | **-** | 99.75 | 99.51[#] |
| | | Jiang et al. [61] | 10.7 | Autoencoder + CNN | 2 | **-** | **-** | **-** | 93.70 | **-** |
| | | Krasteva, V. et al. [62] | 3.4 | Decision Tree | 2 | **-** | **-** | **-** | - | 88.16 |
| | | **Proposed** | - | Self-attentive canonical fusion & CNN | 4 | **450k** | **82.7M** | **34ms** | **96.35** | **96.21**[**] |

[♩]All benchmarks are performed using AMD EPYC 7B12
[#]Shows slightly better F1 scores than rECGnition_v2.0 as the models used fine-graded handcrafted features and were extensively hyper-tuned.
[*]rECGnition_v2.0 was developed/trained for classifying 10 classes.
[**]Transferability test showing rECGnition_v2.0's generalizing capability, with high accuracy and F1 scores on new/diverse datasets.
[##]Model characteristics were determined based on the availability of model architecture given in the research article.
[β]Citation index (citations/year) [63]

The trade-off emphasizes the strategic benefit of rECGnition_v2.0 in situations when quick diagnosis is crucial. In addition, rECGnition_v2.0 demonstrates exceptional model compactness, with just 450,000 trainable parameters. This stands in sharp contrast to other models in the area, which often surpass 1 million parameters. This aspect of rECGnition_v2.0 not only simplifies the process of storing and using it, but also minimizes the model's impact on the environment, in line with the increasing need for sustainable AI solutions.

## 6. Conclusion

A unique deep-learning architecture was proposed to enhance the accuracy of patient-specific diagnosis for individuals with arrhythmia. The rECGnition_v2.0 model shows its enhanced capacity for representing ECG features while also including the understanding of patient context. In comparison to traditional CNN models that have been previously used as ECG feature extractors in research on arrhythmia classification, the rECGnition_v2.0 demonstrates a much lower computational cost. This is supported by the observation that the rECGnition_v2.0 requires considerably fewer floating-point operations (82.7 million FLOPs). Significantly, the computational need of the mentioned model is around 20% of MobileNet with an alpha value of 1 and input dimensions of 128x128x1. Furthermore, it is only 1% of the computational requirement of EfficientNetB0 with input dimensions of 128x128x1. Additionally, rECGnition_v2.0 was endowed with a modest 450,000 trainable parameters. The study's primary contribution was the incorporation of the Attention-based Canonical Correlation layer (SACC), which effectively facilitated the model in identifying individual-specific variations in ECG morphology. The research demonstrated notable prediction results, as supported by an accuracy and F1-score of 98.07% and 98.05%, respectively in a classification task including ten different arrhythmia classes. Significantly, this study not only expands the frontiers of scholarly comprehension but also introduces a concise and widely applicable framework that has great potential for tailored, immediate therapeutic treatments.

**CRediT author statement**

**Shreya Srivastava:** Conceptualization, Methodology, Software, Validation, Formal analysis, Investigation, Data curation, Visualization, Writing - original draft. **Durgesh Kumar:** Software, Validation, Formal analysis, Investigation, Data curation, Visualization. **Ram Jiwari:** Formal analysis, Mathematical Validation, Writing – reviewing and editing. **Sandeep Seth:** Conceptualization, Writing - reviewing and editing. **Deepak Sharma:** Conceptualization, Data curation, Visualization, Writing - reviewing and editing, Resources, Project administration, Funding acquisition, Supervision.

**Acknowledgment**

D.S. acknowledges the financial support from MHRD (BT/2014-15/Plan/P-955 and BIO/FIG/100700), SERB (ECR/2016/001566), DBT (BT/PR40141/BTIS/137/16/2021) and DHR (R.11013/51/2021-GIA/HR and R.11017/36/2023-GIA/HR). S.S. is thankful to MHRD for research fellowship.

# Supplementary Information

**Table S1**
Comparison of validation loss and accuracy for different dropout values for regularization of rECGnition_v2.0.

| Dropout Value | Best Loss | Best Accuracy |
|---|---|---|
| {0.2, 0.1} | 0.08620 | 99.66 |
| {0.2, 0.2} | 0.08545 | 99.62 |
| {0.25, 0.15} | 0.08541 | 99.67 |
| {0.25, 0.25} | 0.08305 | 99.67 |
| {0.25, 0.1} | 0.07863 | 99.89 |

**Table S2**
Performance results for rECGnition_v2.0 (Medium variant) for 10 class arrhythmia classification.

| Class | rECGnition_v2.0 (Medium variant) | | |
|---|---|---|---|
| | P | R | F1 |
| / | 100.0 | 100.0 | 100.0 |
| A | 89.14 | 83.47 | 86.21 |
| F | 89.74 | 79.55 | 84.34 |
| L | 100.0 | 99.53 | 99.76 |
| N | 97.76 | 98.90 | 98.33 |
| R | 99.13 | 98.95 | 99.04 |
| V | 95.91 | 96.45 | 96.18 |
| a | 80.0 | 57.14 | 66.67 |
| f | 100.0 | 100.0 | 100.0 |
| j | 71.43 | 41.67 | 52.63 |

*N*: Normal beat; *L*: Left Bundle Branch Block beat; *R*: Right Bundle Branch Block beat; *V*: Premature Ventricular Contraction beat; */*: Paced beat; *A*: Atrial Premature beat; *f*: Fusion of paced and normal beat; *F*: Fusion of ventricular and normal beat; *j*: Nodal escape beat; *a*: Aberrated atrial premature beat.

**Table S3**

Comparison of performance results for without and with Dual Pathway for 10 class arrhythmia classification.

| Class | Without Dual Pathway | | | With Dual Pathway | | | ΔF1 |
|---|---|---|---|---|---|---|---|
| | P | R | F1 | P | R | F1 | |
| / | 99.74 | 99.74 | 99.74 | 99.74 | 99.74 | 99.74 | 0 |
| A | 92.58 | 86.53 | 89.45 | 92.54 | 86.12 | 89.22 | -0.23 |
| F | 89.04 | 85.53 | 87.25 | 91.04 | 80.26 | 85.31 | -1.94 |
| L | 99.88 | 99.88 | 99.88 | 99.75 | 99.88 | 99.81 | -0.07 |
| N | 98.26 | 99.11 | 98.68 | 98.32 | 99.05 | 98.68 | 0 |
| R | 99.32 | 99.15 | 99.23 | 99.32 | 99.15 | 99.23 | 0 |
| V | 96.63 | 96.49 | 96.56 | 96.78 | 96.91 | 96.85 | 0.29 |
| a | 75.00 | 54.55 | 63.16 | 66.67 | 90.91 | 76.92 | **13.76** |
| f | 100.0 | 95.24 | 97.56 | 100.0 | 100.0 | 100.0 | **2.44** |
| j | 76.92 | 68.97 | 72.73 | 80.77 | 72.41 | 76.36 | **3.63** |

*N*: Normal beat; *L*: Left Bundle Branch Block beat; *R*: Right Bundle Branch Block beat; *V*: Premature Ventricular Contraction beat; */*: Paced beat; *A*: Atrial Premature beat; *f*: Fusion of paced and normal beat; *F*: Fusion of ventricular and normal beat; *j*: Nodal escape beat; *a*: Aberrated atrial premature beat.

**Table S4**

Comparison of CNN-based state-of-the-art model with rECGnition_v2.0 for model parameters, FLOPs, and Accuracy[*].

| Model | Parameters | FLOPs | Memory requirement for prediction | Prediction time per sample | Accuracy |
|---|---|---|---|---|---|
| MobileNet | 3.2M | 366M | 44MB | 0.21ms | 97.0 |
| EfficientNetB0 | 4M | 252M | 57MB | 0.38ms | 97.0 |
| Resnet50 | 23.5M | 2.47B | 220MB | 0.55ms | 96.0 |
| rECGnition_v2.0 | **450k** | **82.7M** | **15MB** | **0.16ms** | **98.0** |

[*]All benchmarks were performed using Google v3-8 TPU

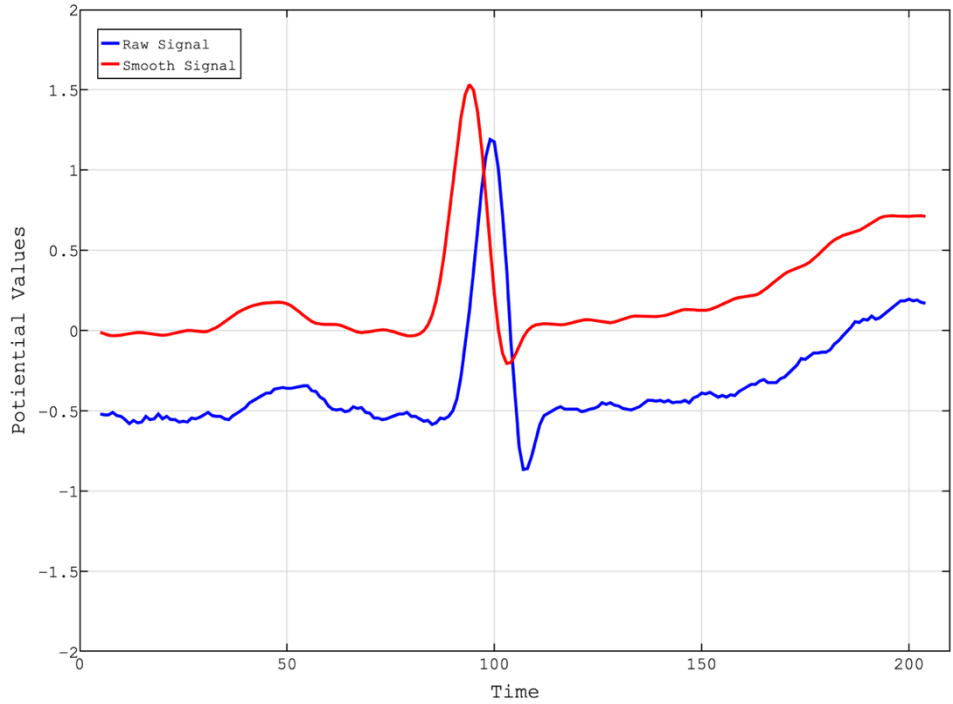

**Figure S1**: Comparison between raw ECG beat segmented from MITDB and processed beat after applying smoothening and baseline correction.

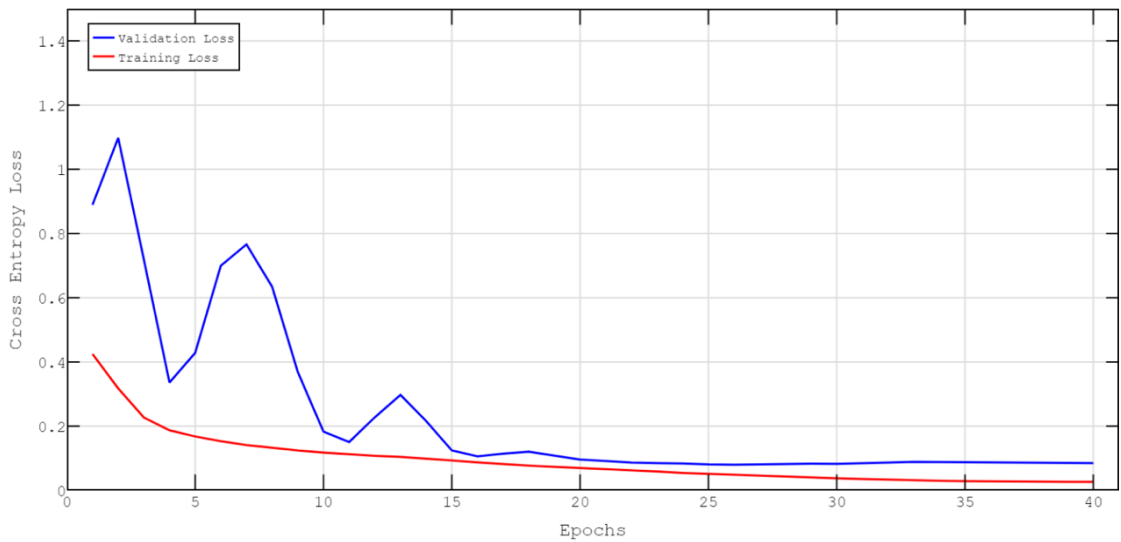

**Figure S2**: Training and validation loss convergence plot.

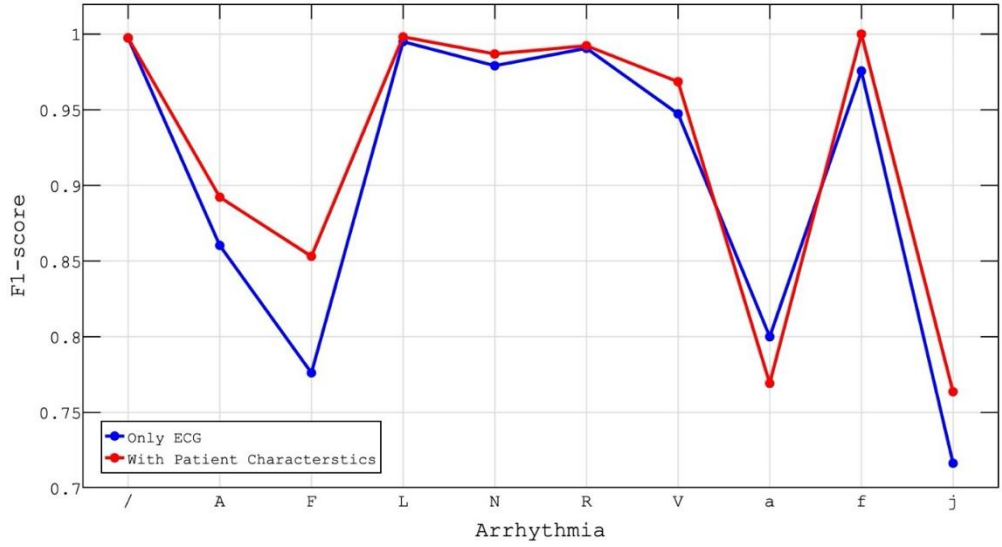

**Figure S3**: Comparison of F1-score using only ECG heartbeat image and using Patient characteristics data along with ECG heartbeat image.

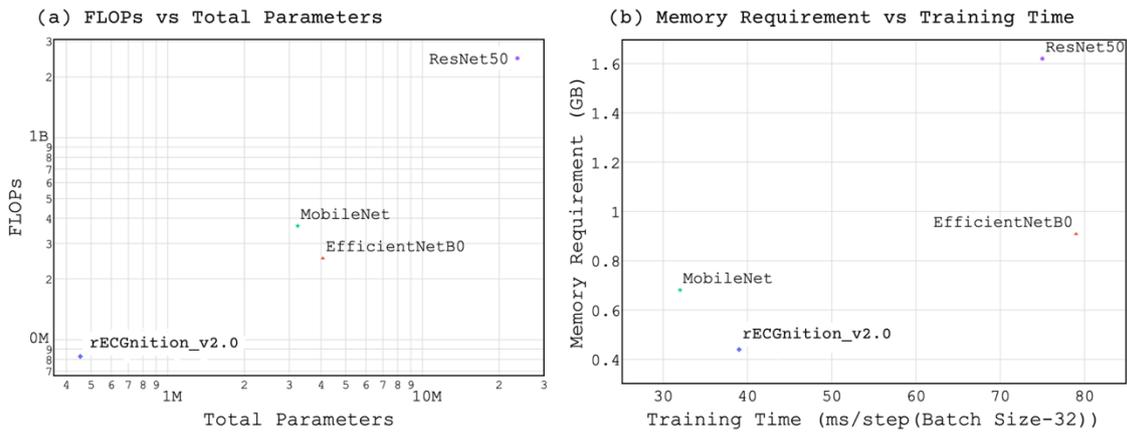

**Figure S4**: Comparison of computational requirements and efficiency of MobileNet, EfficientNetB0, Resnet50, and rECGnition_v2.0.

27